\documentclass[sigconf, nonacm]{acmart}
\usepackage{bytefield}
\usepackage{url}

\newcommand\vldbdoi{XX.XX/XXX.XX}

\newcommand\vldbavailabilityurl{https://github.com/lancedb/lance-research/tree/main/file_2_1}

\begin{document}
\title{Lance: Efficient Random Access in Columnar Storage through Adaptive Structural Encodings}

\author{Weston Pace}
\affiliation{%
  \institution{LanceDB}
}
\author{Chang She}
\affiliation{%
  \institution{LanceDB}
}
\author{Lei Xu}
\affiliation{%
  \institution{LanceDB}
}
\author{Will Jones}
\affiliation{%
  \institution{LanceDB}
}
\author{Albert Lockett}
\affiliation{%
  \institution{LanceDB}
}
\author{Jun Wang}
\affiliation{%
  \institution{LanceDB}
}
\additionalaffiliation{
  \institution{Clark University}
}
\author{Raunak Shah}
\affiliation{%
  \institution{LanceDB}
}
\additionalaffiliation{%
  \institution{University of Illinois, Urbana-Champaign}
}

\begin{abstract}
The growing interest in artificial intelligence has created workloads that require both sequential and random access.  At the same time, NVMe-backed storage solutions have emerged, providing caching capability for large columnar datasets in cloud storage.  Current columnar storage libraries fall short of effectively utilizing an NVMe device's capabilities, especially when it comes to random access.  Historically, this has been assumed an implicit weakness in columnar storage formats, but this has not been sufficiently explored.  In this paper, we examine the effectiveness of popular columnar formats such as Apache Arrow, Apache Parquet, and Lance in both random access and full scan tasks against NVMe storage.

We argue that effective encoding of a column's structure, such as the repetition and validity information, is the key to unlocking the disk's performance.  We show that Parquet, when configured correctly, can achieve over 60x better random access performance than default settings. We also show that this high random access performance requires making minor trade-offs in scan performance and RAM utilization.  We then describe the Lance structural encoding scheme, which alternates between two different structural encodings based on data width, and achieves better random access performance without making trade-offs in scan performance or RAM utilization.
\end{abstract}

\maketitle

\begingroup
\renewcommand\thefootnote{}\footnote{\noindent
This work is licensed under the Creative Commons BY-NC-ND 4.0 International License. Visit \url{https://creativecommons.org/licenses/by-nc-nd/4.0/} to view a copy of this license. For any use beyond those covered by this license, obtain permission by emailing \href{mailto:contact@lancedb.com}{contact@lancedb.com}. Copyright is held by the owner/author(s). 
}\addtocounter{footnote}{-1}\endgroup


\section{Introduction} \label{introduction}

In recent years, a wide array of data lakes, data warehouses, and similar solutions have emerged to target an equally diverse set of data storage use cases\cite{DataHouseReview}.  Despite their differences, these systems share some common design principles.  First, they utilize columnar representations, both in-memory and in storage, to achieve efficient compression and vectorized compute\cite{ColumnStores06}.  Second, they store data in cloud storage\cite{CloudStorage}.  Cloud storage can provide extremely high bandwidth but supports a relatively low number of operations per second compared to local NVMe-based storage.  Finally, many of these systems are built on top of shared, composable, open source components\cite{ComposableData}.

These systems perform especially well on full scan analytics workloads\cite{Datafusion}.  This is essential for machine learning tasks such as model training, feature extraction, and scientific analysis.  These workflows can require tens of thousands of columns, and often columns have significant struct or list nesting\cite{Bullion}.  Columnar file formats, such as Apache Parquet\cite{Parquet} and Apache Orc\cite{Orc}, are well suited for this task and new formats such as Nimble\cite{Nimble}, Vortex\cite{Vortex}, and Lance\cite{Lance2} are emerging to address weaknesses in these formats.

However, columnar storage systems have historically performed poorly on search-oriented workflows such as full-text search,  semantic search, and retrieval-augmented generation\cite{ColumnarEvaluation}.  These search workflows are equally important for the inference and training paths in many machine learning applications.  Search workloads typically need to fetch small subsets of results that are not aligned along a primary (i.e. clustered) index and, as a result, they require large amounts of random access I/O operations (IOPS).  This is a challenge for both columnar file formats and cloud storage.

\begin{figure}[h]
    \centering
    \includegraphics[width=8cm]{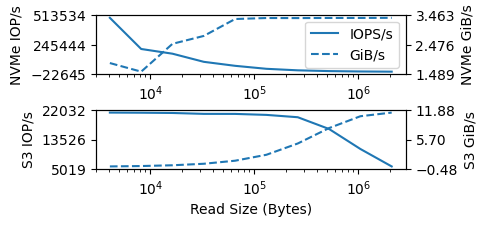}
    \Description{A chart showing how random access and full scan performance fluctuate with I/O request size in different storage solutions.  Small requests achieve good random access performance but large requests are required to achieve good full scan performance.   The trade-off point depends on the storage solution.  NVMe can handle very small requests while cloud storage cannot.}
    \caption{NVMe storage introduces unique performance characteristics}
    \label{fig:disk-profile}
\end{figure}

Figure \ref{fig:disk-profile} demonstrates this challenge for cloud storage.  We perform a simple file benchmark comparing data access between an NVMe disk (Samsung 970 EVO Plus) and S3 (c7gn.8xlarge instance).  We measure bandwidth with sequential reads of different sizes and measure random access with random reads of different sizes.  Services like S3 are generally limited to tens of thousands of IOPS \cite{S3PerformanceGuidelines} and do not benefit from reads smaller than about 100KB.  NVMe-based storage can support hundreds of thousands of IOPS and requires disk-sector sized reads of 4KiB to achieve maximum random access throughput.

One historical way to address this cloud storage problem has been to keep a copy of the data in-memory \cite{PredicateCaching}, but this is expensive.  A more affordable solution is to utilize a caching layer based on NVMe storage \cite{NvmeCaching}.  This is especially appealing for modern approaches based on table formats utilizing multi-version concurrency control, since data files are read-only and easily cached.  New services have emerged to provide this caching service such as WekaFS \cite{WekaFs} and JuiceFS \cite{JuiceFs}.  Alternatively, services like S3 Express can provide low latency cloud storage, potentially eliminating the need for a separate cache.

These new storage solutions require a reevaluation of the file formats used to store data.  To date, there has been limited benchmarking of columnar storage for search-based workflows.  \cite{ColumnarEvaluation} surveys ORC and Parquet on a vector search task.  In \cite{Rottnest} it was shown that Parquet can be used for vector search against cloud storage, effectively saturating the limited IOPS available in S3.  This paper expands on those endeavors to provide a more thorough and detailed analysis, focused on NVMe storage.  In addition, we introduce and describe Lance, a new columnar format inspired by this analysis.

\subsection{Overview}

Columnar file formats store tabular data in a column-major order.  This means row-based random access requires at least one or more IOPS per column.  However, we find that this problem is potentially made worse by the way that compression and nested structure are encoded.  As a result, file formats can require many IOPS per column, require IOPS with a large amount of read amplification (reading more data than strictly required), or require a large amount of compute overhead per access.

We will first describe the general structure of columnar storage formats.  We then explain, in detail, the methods used to structurally encode compressed buffers and nesting information.  We will show how this structural encoding impacts random access performance and forces existing formats into trade-offs.  We introduce the Lance format and describe how its adaptive structural encoding avoids these trade-offs.  Finally, we benchmark existing formats and the Lance format to verify our assumptions.

\section{Columnar Storage}

Columnar storage describes how tabular data is encoded into a binary file.  The file format itself handles the overall structure of the file and major components.  Within the format, there are structural encodings which control how nested data types are split into individual arrays or buffers.

To aid understanding, we examine several common file formats and compare and contrast the approaches they take.  The Apache Parquet format is one of the most ubiquitous columnar formats in use today.  It aims to maximize columnar compression and scanning performance.  The format is not very clearly versioned, as different implementations support different sets of optional features.  This paper primarily focuses on the features offered by the parquet-rs implementation as of version 54.  In particular, it has a good implementation of the page offset lookup structure \cite{PqPageOffset}, which is critical for random access.

The Apache Arrow IPC format\cite{ArrowFile} describes how to encode Arrow's in-memory format into a file with minimal copying or modification of data.  General buffer compression is an optional feature, but the choices of compression and encoding are not as sophisticated as the other formats.  We examine the 1.5 version of the format, which is the latest available at the time of writing.

The Lance format is a new format developed by the authors of this paper.  The aim is to achieve a balance between full scan and random access workloads.  This paper studies the 2.1 version of the format, which is still considered experimental.  The 2.0 format is also used in our experiments, as it utilizes a structural encoding similar to Arrow and provides a way to test that encoding's potential for random access reads.

\subsection{File Format}

\begin{figure*}
    \centering
    \includegraphics[width=18cm]{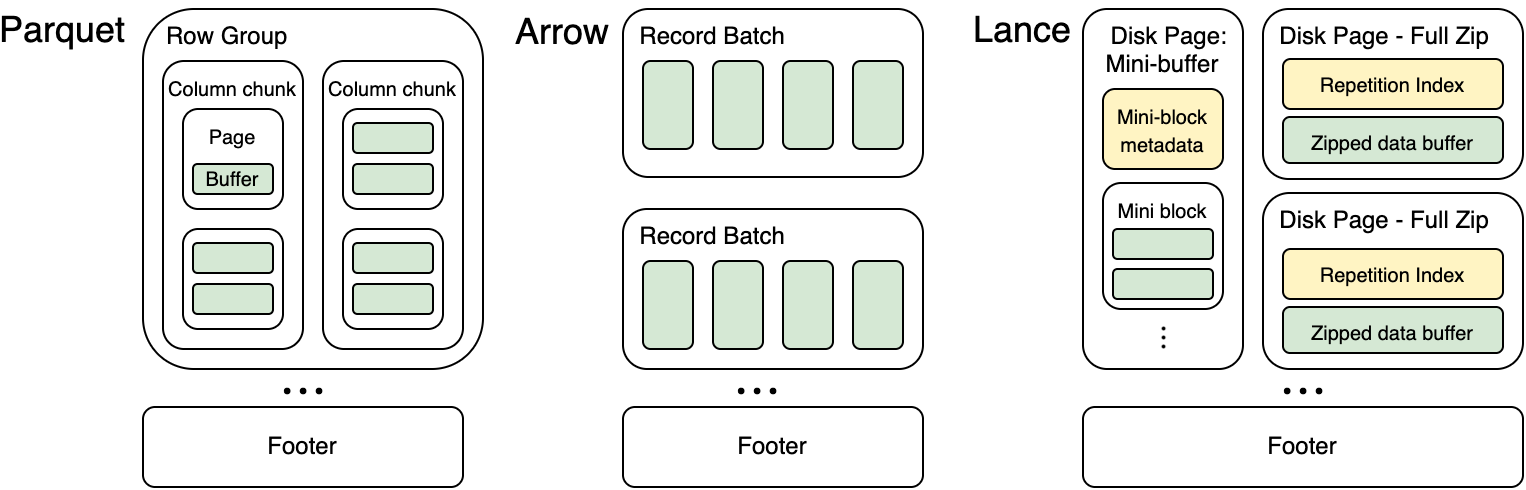}
    \Description{A diagram showing an overview of the file formats.  It shows the different components described in section 2.1}
    \caption{An overview of the major components of each of the file formats}
    \label{fig:encodings-overview}
\end{figure*}

The file format describes the major components of the file.  All of the formats we examine have similar components.  We will briefly describe them here to standardize terminology.  In the rest of this paper, we will use the Parquet terminology regardless of the format being examined.  Figure \ref{fig:encodings-overview} gives an overview of the different formats.

A \textbf{row group} represents a block of tabular data within the file.  A file can consist of one or more row groups.  Row groups were originally designed to allow for performing a full scan of the file in parallel.  In Arrow, the concept is called a \textit{record batch}.  Lance does not have an equivalent concept and achieves parallel scans without them, as detailed in \cite{Lance2}.  A Lance file can be thought of as always having a single row group.

A \textbf{column chunk} represents a collection of buffers, stored contiguously, belonging to a single column of data.  In Arrow, there is no equivalent component (all buffers belong to the record batch).  In Lance, the concept is called a \textit{disk page} (not to be confused with Parquet's concept of a page).  In Parquet, each column in the schema has one column chunk per row group.  In Lance, a single column is allowed to have multiple column chunks throughout the file.  Row groups and column chunks are not important for random access performance, but they are critical for full scan performance.

A \textbf{page} is a small section of a column chunk.  In Parquet, pages are utilized to enhance the effectiveness of compression (most compression techniques benefit from some level of chunking).  The exact contents of a page will depend on the data type and encoding.  For example, a page of binary data will contain a buffer of offsets and a buffer of data.  Pages also contain buffers for repetition and definition levels.  Repetition and definition levels were originally described in \cite{Dremel} and we find them to be important for random access performance as they are an effective way to combine multiple buffers of offsets and validity.  Pages also can contain additional metadata, such as statistics or bloom filters, to assist in pushdown filtering.

In the Arrow IPC format there is no concept equivalent to a page.  Arrow has no formal spec for statistics and the only supported compression mechanisms operate on the entire buffer.  Lance 2.0 has no concept equivalent to Parquet's page.  Compression is limited and not chunked.  In Lance 2.1 the presence of pages will depend on the structural encoding used.  In the miniblock encoding (described later) Lance 2.1 will utilize a "miniblock chunk" which is roughly equivalent to Parquet's page.  Pages are a crucial factor in random access performance as they introduce and control read amplification, and this will be discussed in more detail.

\subsection{Compression}

This paper is not a detailed examination of compression techniques.  However, we have found that the compression approaches available depend on the structural encoding, and so some discussion is warranted.  We categorize compression encodings in a few different ways.

\textbf{Transparent} encodings compress values without introducing dependencies between data values.  If the compression is transparent, then a single value can be extracted from a compressed buffer if the location and length of that value are known.  Examples of transparent encodings include bit packing, FSST, and dictionary encoding.  By contrast, \textbf{opaque} encodings do introduce dependencies between data values.  Decompression of a single value is impossible, and multiple values must be decompressed to access a single value.  Examples of opaque encodings include back-referencing encodings such as Snappy and delta-based encodings such as Parquet's delta-length byte array encoding.

An opaque encoding can be used in a transparent fashion if it is applied on a per-value basis.  For example, for very large values, Lance will apply LZ4 compression on a per-value basis.  Each value is an independent LZ4 frame and can be decompressed by itself.

We define the \textbf{sparse} compression of data as storing the data with nulls removed.  This is often possible because the nullability is stored elsewhere (e.g. in a validity bitmap) and can be restored after the fact.  However, this makes it more difficult to access the $i_{th}$ value even if the items have fixed size.  The counterpart, \textbf{dense} compression, encodes filler data in the place of nulls, enabling random access but requiring more space.

\subsection{Evaluation Criteria}

In the next section we will discuss structural encodings in depth.  When considering random access performance we are chiefly interested in the number of IOPS that must be performed to reach a single value and the read amplification present in those IOPS.

We make a special exception for metadata that we will refer to as the \textbf{search cache}.  We define the search cache as small (we aim for 0.1\% of the data size) auxiliary data which is read when a file is opened (or first searched) and cached in memory.  The I/O cost can be ignored because the metadata is loaded into memory once and then cached for many searches against the data.  The cost of loading the data is amortized among these searches and becomes negligible.  In our evaluation, we only consider warm searches where the metadata is already in memory.  This more accurately models the real-world search use cases we have encountered.

When measuring full scan performance we are chiefly considering compression and the total amount of data that must be read from disk.  Since full scans read entire column chunks in a single IOP, the number of IOPS is rarely a concern.  The search cache is typically \textit{not} cached in memory in full-scan applications and may not need to be read from disk at all.  In our full-scan results, we do not assume that any metadata has been cached in memory.

\section{Columnar Encoding Strategies} \label{strategies}

In this paper, we identify two stages of columnar encoding.  \textbf{Structural encoding} is the first step of columnar encoding and converts a potentially nested array into one or more leaf nodes, which each contain one or more primitive arrays and/or buffers.  This process is sometimes called \textit{shredding}.  \textbf{Compressive encoding} then takes each of these leaf arrays or buffers and compresses them into the smallest possible representation.

There has been much discussion on compressive encodings.  FastLanes\cite{FastLanes} describes effective strategies for integer compression, ALP\cite{Alp} describes floating-point compression, and BtrBlocks\cite{BtrBlocks} explains how encodings can generally be cascaded together.

There has been less discussion on structural encoding techniques.  These are often implicitly defined by the file format and not explored thoroughly or well understood.  We have found structural encoding to be crucial to achieving effective random access.

Structural encodings define how a column chunk is converted into one or more buffers to store on the disk.  Different structural encoding techniques may convert an input array into a different number of buffers.  In addition, the choice of structural encoding impacts which compression techniques can be used.  Finally, structural encodings determine how much read amplification and what sort of scheduling must be done to read the data.

We will now describe the structural encodings employed by various file formats in detail.  In these descriptions, we will refer to buffers $b_0, b_1, ...$ which represent contiguous slices of data.  We use $b_i[x]$ to denote the item at offset $x$ from the buffer $b_i$.  This might be the read of a single bit (e.g. validity buffer), byte (binary data), or several bytes (e.g. string length or i64 value).  We use $b_i[x..y]$ to denote the range of values starting at offset $x$ and leading up to (but not including) offset $y$.

\subsection{Parquet Structural Encoding Scheme}

Parquet converts a nested array into a series of flattened leaf columns.  Each leaf column contains a flattened list of items (with nulls removed), a buffer of repetition levels, and a buffer of definition levels.  The flattened list of items is compressed using a chunked compression strategy.  Chunks are referred to as pages.

\begin{figure}[h]
    \centering
    \begin{bytefield}[bitheight=3em]{40}
      \bitbox{10}{rep$[0..p0]$} & \bitbox{10}{def$[0..p0]$} & \bitbox{10}{$b_0[0..p0]$} & \bitbox{10}{$b_1[0..p0]$} \\
      \bitbox{10}{rep$[p0..p0+p1]$} & \bitbox{10}{def$[p0..p0+p1]$} & \bitbox{10}{$b_0[p0..p0+p1]$} & \bitbox{10}{$b_1[p0..p0+p1]$} \\
      \wordbox[]{1}{$\vdots$} \\[1ex]
    \end{bytefield}
    \Description{The image contains two boxes stacked on top of each other.  The top box is split into four boxes.  The first contains the repetition buffer from 0 to p0.  The second contains the definition buffer from 0 to p0.  The third contains the lengths buffer from 0 to p0.  The fourth contains the data buffer from 0 to p0.  The lower box is split into four similar boxes except they range from p0 to p0 + p1. }
    \caption{An interpretation of a column chunk containing binary data (delta-length binary packed encoding) which has a buffer of lengths ($b_0$) and data ($b_1$) as a series of pages with $p0, p1, ...$ values in each page}
\end{figure}

Each page contains the relevant repetition and definition levels for the values on the page.  The exact number of buffers is irrelevant as the page is considered an opaque chunk and is not divisible.  This allows for sparsely encoding values (null items occupy no space in the items buffer) and the use of opaque compression algorithms such as delta encodings.

Some Parquet readers and writers also support a page offset index.  The page offset index is what we use as our search cache.  This structure contains the location of each page and the index of the first row in the page.  By performing a binary search into the page offset index we can identify exactly which page we need to load for a given row index.  This scheme does introduce read amplification equal to the page size.  However, if the page size is made small enough, then Parquet can perform efficient random access.

Parquet's encoding scheme is ideal for random access in many cases.  The page offset index allows for quick identification of pages that must be loaded, and only a single IOP is required for each page.  There are, however, some potential issues.  First, Parquet's expensive compression may keep the CPU busy, which makes it difficult to keep the disk queue full.  The second is that the page offset index requires too much RAM for large data types such as vectors (we discuss this more in section \ref{search-cache}.)

\subsection{Arrow Structural Encoding Scheme} \label{arrow-structural}

Arrow also converts nested data into a series of flattened leaf columns.  However, these leaf columns are dense.  Null values must be represented by placeholder \textit{garbage bytes}.  Repetition and definition levels are not employed.  Instead, each layer of repetition has an array of offsets.  Each layer of definition has a bitmap of validity.

There is no concept of pages or chunking.  The only compression techniques available are general compression that cover the entire column chunk and render the entire column chunk opaque.  It is common to store Arrow files without any compression.

These properties are well suited for random access against in-memory storage which is not typically throttled by IOPS/second.  If the data is uncompressed, then there is no read amplification and no CPU decode cost.  Furthermore, there is no need for a page offset index and no search cache.

However, we find the Arrow structural encoding to be poorly suited for disk or cloud storage when the data type has multiple buffers.  This can happen with nested types like struct and list, and with variable width types like strings.  We must perform more IOPS than Parquet because we are loading data from multiple buffers.

For example, a \texttt{List<String>} array, which contains nulls in each layer, will require 5 IOPS per value.  These IOPS will need to be issued in 3 phases.  First, the list validity and list offsets are read.  Then, the string validity and string offsets are read.  Finally, the string values are read.  These IOPS are unlikely to be coalesced unless the row group (record batch) is extremely small.  However, extremely small row groups are ineffective for full scan columnar access.

\begin{figure}[h]
    \centering
    \begin{bytefield}{32}
      \wordbox{1}{List validity ($b_0[10]$)} \\
      \wordbox{1}{List offsets ($b_1[10:12]$) }\\
      \wordbox{1}{String validity ($b_2[b_1[10]:b_1[11]]$)} \\
      \wordbox{1}{String offsets ($b_3[b_1[10]:b_1[11]]$)} \\
      \wordbox{1}{String data ($b_4[b_3[b_1[10]]:b_3[b_1[11]]]$)} \\
    \end{bytefield}
    \Description{This images shows 5 boxes.  The first access the list validity buffer, b0, at location 10.  The second accesses the list offsets bufffer, b1, at locations 10 and 11.  The third accesses the string validity buffer, b2, based on the offsets from b1.  The fourth accesses the string offsets buffer, b3, based on the offsets from b1.  The final box accesses the string data buffer, b4, based on the offsets from b3.}
    \caption{An example showing the layout and accesses needed to fetch the 10th item in a \texttt{List<String>} array.}
\end{figure}

Even though Arrow only describes general opaque compression, there are transparent compressive encodings that could be used with the Arrow structural encoding.  For example, the list and string offsets could be bit packed with the bit width stored in the metadata, and we would still be able to perform random access.  With simple types, we can thus get a file that has good random access and full scan characteristics.  This was the approach we planned for in the 2.0 version of Lance before we realized an excessive number of IOPS would still be required for complex types.  In this paper, we only briefly evaluate Arrow-based approaches, as this limitation renders them unsuitable for our uses.

\section{Lance Structural Encoding Scheme} \label{lance-structural}

Lance's structural encoding scheme was designed to provide effective random access performance in all situations, while keeping the search cache small.  Specifically, we have the following goals:

\begin{itemize}
    \item At most 1 IOP for random access to a fixed-width column
    \item At most 2 IOPS for random access to a variable-width column
    \item Performance should be consistent regardless of how many levels of nesting there are, the size of the data type, or the compression used.
    \item Full scan performance should be comparable with Parquet
\end{itemize}

At a conceptual level, this is achieved by taking the various buffers that make up a column and zipping them together.  This conceptual idea is followed very closely in large data types where we utilize the \textbf{full zip} structural encoding.  For smaller data types, this zipping process becomes too computationally expensive and we must utilize the \textbf{miniblock} structural encoding, which is more similar to Parquet.  Both encoding strategies utilize repetition and definition levels.  These are constructed in roughly the same manner as Parquet.

\subsection{Full Zip Encoding}

The full zip encoding strategy is suitable for large data types.  We use 128 bytes per value as a threshold based on experimental measurements.  This includes types such as vector embeddings, tensors, images, and large text fields.  Because these types are large, we can afford to spend more effort per value without impacting scan performance.  In addition, these types are often already compressed (such as vector embeddings or compressed images) or do not require large chunks to gain effective compression.

\begin{figure}[h]
    \centering
    \begin{bytefield}{32}
      \bitbox{6}{$rep[0]$} & \bitbox{6}{$def[0]$} & \bitbox{10}{$b_0[0]$} & \bitbox{10}{$b_1[0]$} \\
      \bitbox{6}{$rep[1]$} & \bitbox{6}{$def[1]$} & \bitbox{10}{$b_0[1]$} & \bitbox{10}{$b_1[1]$} \\
      \wordbox[]{1}{$\vdots$} \\[1ex]
    \end{bytefield}
    \Description{This images shows two boxes stacked on top of each other.  The first box contains repetition at 0, definition at 0, b0 at 0, and then b1 at 0.  The second box contains repetition at 1, definition at 1, b0 at 1, and then b1 at 1.}
    \caption{The full-zip encoding transposes the buffers into a row-major order.}
\end{figure}

In this strategy, the repetition levels, definition levels, and all buffers that make up the leaf column are zipped into a single buffer of values.  Note that only the buffers in a single primitive leaf column are zipped together.  Zipping together multiple columns (such as multiple columns within a struct) is a different concept we refer to as \textit{packing} which is discussed later.  When the data type has a fixed-width, then the zipped buffer will also have a fixed-width.

\subsubsection{Repetition and Definition Levels}

Repetition and definition levels are first bit packed and combined into a 1-4 byte word that is referred to as a control word.  We do not employ chunking or run length encoding when we bit-pack the repetition and definition levels.  As a result, the byte width of the control word is consistent throughout the column chunk.  Every value is then preceded by a control word containing its repetition and definition information.  Repetition indices (described later) point to the control word as the start of a value.

For example, a \texttt{Struct<List<String>{}>} will require up to 3 bits for definition level and 1 bit for repetition level.  As a result, the control word will be 1 byte per value.  Control words themselves are not bit packed as the compression ratio is marginal (less than 1.01x since we know the data type is at least 128 bytes and our savings will be less than 1 byte) and it would introduce non-byte-aligned values.  It is quite rare for a control word to require more than 1 byte.

\subsubsection{Data Buffers}

Simple fixed-width layouts are already a single buffer and no zipping is required.  Given variable-width layouts, we first convert offset buffers into a buffer of lengths, and this length is placed before the value.  Lengths are bit-packed to the nearest byte boundary and are, at most, 8 bytes.

\begin{figure}[h]
    \begin{bytefield}{32}
      \wordbox{1}{0bXXXX1000 (new list, valid)} \\
      \wordbox{1}{2} \\
      \wordbox{1}{'AB'} \\
      \wordbox{1}{0bXXXX0000 (same list, valid)} \\
      \wordbox{1}{1} \\
      \wordbox{1}{'C'} \\
      \wordbox{1}{0bXXXX1011 (null list)}  \\
      \wordbox{1}{0bXXXX1100 (null struct)} \\
      \wordbox{1}{0bXXXX1001 (new list, null)} \\
      \wordbox{1}{0bXXXX1010 (new list, empty)} \\
    \end{bytefield}
    \Description{This image shows 10 boxes.  The top box contains "0bXXXX1000 (new list, valid)".  The next box contains 2.  The next box contains 'AB'.  The next box contains "0bXXXX0000 (same list, valid)".  The next box contains 1.  The next box contains 'C'.  The next box contains "0bXXXX1011 (null list)".  The next box contains "0bXXXX1100 (null struct)".  The next box contains "0bXXXX1001 (new list, null)".  The final box contains "0bXXXX1010 (new list, empty)".}
    \caption{The full-zip encoding of \texttt{["AB", "C"], NULL LIST, NULL STRUCT, [NULL], []} with the data type \texttt{Struct<List<String>{}>}}
    \label{full-zip-example}
\end{figure}

An example is provided in figure \ref{full-zip-example}.  There are 3 bits of definition and 1 bit of repetition, and so we have 1-byte control words.  The least significant 3 bits represent the definition level.  $000$ is a valid item, $001$ is a null item, $010$ is an empty list, $011$ is a null list, and $100$ is a null struct.  The next bit indicates the repetition level.  A $1$ indicates a new list, and a $0$ continues the existing list.

\subsubsection{Compression}

It's important to note that values are compressed \textit{before} we zip them together.  This ensures that we get the benefits of columnar compression.  For example, to compress strings, we can apply FSST to the strings and then bitpack the lengths.  We place the symbol table into the metadata for the disk page.  This layout requires that the compressive encodings we use are transparent because we must know which bytes correspond to each value and be able to decompress the value independently.

We also cannot sparsely encode null values when working with fixed length types.  For example, if we are encoding 4KiB vector embeddings and we have a null value, then we must put in 4KiB of filler data after the control word indicating a null value.  This allows us to map an index to a byte range for random access.  For variable length data, we can encode nulls as a control word only.

\subsubsection{Repetition Index}

Without additional information, it is impossible to perform random access to variable-width values encoded with the full zip encoding.  This problem is solved by the \textbf{repetition index}.  The repetition index is a single array of buffer offsets which is bit-packed (without chunking) and stored before the zipped buffer.  The repetition index does not need to be read from disk when performing a full scan of the column.

When performing random access we first look up the position of the value in the repetition index and subsequently look up the value itself.  If we need to read a range of values then we can either read a range from the repetition index (1 IOP) or read the start and end of the range from the repetition index (2 IOPS run in parallel).  This allows random access with 2 IOPS regardless of the nested structure.  In this paper, the repetition index is not part of the search cache, as it is likely too large for high-scale datasets.

\begin{figure}
    \centering
    \includegraphics[width=8cm]{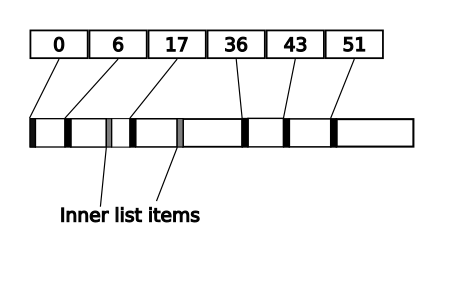}
    \Description{This image shows two buffers.  The top buffer is the repetition index and contains values, 0, 6, 17, 36, 43, 51.  These are byte offsets and each value has a line going to the second buffer at that location.  The second buffer is marked up with a number of black and gray bars.  These bars are control words that start a variable sized value.  The black control words are top-level rows and they each have a line connecting to the repetition buffer.  The gray control words are inner list items and have no connection to the repetition index.  They are labeled in the image "inner list items"}
    \caption{A repetition index points to the control word of each top-level item and enables random access.}
\end{figure}

\subsubsection{Run Length Encoding}

It is possible for full zip encoded data to utilize run-length encoding.  In this case, the repetition index contains three numbers for each value.  The first two are the buffer offset and size (omitted if the data type is fixed-width) and the second is the run offset.  This allows us to read a single value (or range of values) from a run-length encoded column in 2 IOPS.

This concept is similar to the skip tables discussed in \cite{Procella}.  The repetition index does increase the on-disk size of the column.  With run-length encoded data, it is even possible that the repetition index is considerably larger than the zipped buffer.  However, the repetition index is never read on a full scan of the data, and so it has no impact on full scan performance, though it may have some marginal impact on the overall cost of the storage.  We have not yet implemented this in Lance 2.1 and it is not examined in detail here.

\subsection{Mini-block Encoding}

The mini-block encoding complements the full zip encoding and is suitable for small data types.  These data types are small enough that we can accept some amount of read amplification to reduce the amount of compute work necessary to process the data during scans.  In the mini-block encoding, we first divide an array into a series of small chunks.  The goal is for each chunk to represent 1-2 disk sectors (4KiB-8KiB) of compressed data.

Each compressed chunk contains a buffer of repetition levels, a buffer of definition levels, and any number of data buffers.  Since these buffers are not transposed, the encoding and decoding processes are vectorized and more efficient.  In addition, we do not need to store null data, compression can be opaque, and compression is chunked.  However, we must decode an entire chunk of data to retrieve any single value, which introduces both read amplification and, on the random access path, compute amplification.

This approach is quite similar to Parquet although there are a few minor differences.  The page overhead is slightly smaller.  Buffers are required to be 8-byte aligned.  We treat primitive fixed-size-list arrays as primitive types (the validity is another buffer and not part of the repetition or definition).  We also do not require that each chunk begins with a top-level record.

\subsubsection{Chunking}

The exact protocol for choosing chunk sizes and recording chunk metadata is somewhat heuristic.  Our current approach is to require each chunk to have a power-of-two number of items and pad to the next multiple of 8 so that each chunk is 8-byte aligned.  We can then record the chunk size in two bytes.  12 bits record the number of 8-byte words in the chunk.  4 bits record the log base 2 of the number of values.  We further restrict chunks to contain at most 4096 values.  This keeps the search cache small (2 bytes per chunk) and avoids read amplification (miniblocks are at most 32 KiB and usually much closer to 4-8 KiB).

However, there are some disadvantages.  Highly compressible data may be forced into more chunks than necessary.  Compression algorithms are responsible for choosing chunk sizes, which adds complexity to compression implementation.  Chunks are not cleanly aligned to disk sectors.  Finally, picking an appropriate chunk size when there are multiple columns is quite difficult.  We think iterating on this structure may be an interesting area for future research.

\subsubsection{Chunk Structure}

Each chunk has a brief header which records the minimum amount of information required to decode the chunk.  After the header, the buffers are placed so that they start on 8-byte aligned addresses.

\begin{itemize}
    \item the number of buffers in the chunk (2 bytes)
    \item the size of each buffer (2 bytes per buffer)
    \item for each buffer:
    \begin{itemize}
        \item padding to 8-byte aligned value
        \item buffer data
    \end{itemize}
    \item padding to 8-byte aligned value
\end{itemize}

\begin{figure}[h]
    \begin{bytefield}{32}
      \bitbox{8}{3} & \bitbox{8}{31} & \bitbox{8}{248} & \bitbox{8}{4957} \\
      \wordbox{1}{$b_0$ (def levels)} \\
      \wordbox{1}{...} \\
      \bitbox{28}{$b_0$} & \bitbox{4}{XX} \\
      \wordbox{1}{$b_1$ (string offsets)} \\
      \wordbox{1}{...} \\
      \wordbox{1}{$b_1$} \\
      \wordbox{1}{$b_2$ (string data)} \\
      \wordbox{1}{...} \\
      \bitbox{20}{$b_2$} & \bitbox{12}{XX XX XX} \\
    \end{bytefield}
    \Description{This image contains 10 boxes stacked on top of each other.  The first contains the values 3, 31, 248, and 4957.  The second contains "b0 (def levels)".  The second contains "..."  The third contains "b0" followed by "XX" (padding).  The fifth contains "b1 (string offsets)".  The sixth contains "..."  The seventh contains "b1".  The eighth contains "b2 (string data)."  The ninth contains "..."  The tenth contains "b2" followed by "XX XX XX (padding)".}
    \caption{Example miniblock chunk containing 248 nullable strings with 4,957 bytes of data}
    \label{mini-block-example}
\end{figure}

\subsubsection{Repetition Index}

When the array has one or more variable-size list arrays, then we need a repetition index to determine which chunks need to be read when performing random access.  This is necessary because the chunk metadata only contains the number of flattened items and does not contain the number of complete rows.  In addition, some complexity is introduced by the fact that rows may be split across chunks.

The repetition index is stored on a per-chunk basis and contains N + 1 non-negative integers where N is the maximum level of random access nesting that is desired.  For example, to support lookups such as \texttt{array[x]} we need 2 values per chunk.  To support lookups such as \texttt{array[x][y][z]} we need 4 values per chunk.  The first value is the number of top-level lists (i.e. rows) completed since the start of the chunk.  The second value is the number of 2nd level lists completed since the end of the last top-level list and so on.  The final value is the number of flattened items since the end of the last X-th level list where X is the nesting level we are supporting.  This design allows us to handle an arbitrary depth of nested lookups.  However, the Lance 2.1 reader and writer currently support only a single level of list lookup.

\subsubsection{Search Cache} \label{search-cache}

The Parquet file format has only a single encoding scheme.  If we use large values and small page sizes we end up with one page per value.  The in-memory size of the offset index in parquet-rs is currently 20 bytes per page.  As a result, we would need 20 GiB of search cache per billion rows for these large columns.

In Lance, our search cache depends on the structural encoding.  The full zip encoding does not have a search cache.  This allows us to overcome Parquet's main difficulty, which is dealing with large data types.  With small data types, we use the miniblock encoding which requires that we know the miniblock metadata, as well as any repetition index. Also, compressive encodings may store auxiliary data in the search cache such as dictionaries or symbol tables.

Currently, Lance requires 24 bytes per chunk without a repetition index and 41 bytes per chunk with a repetition index.  Even so, we only use this encoding for small data types, and we will always have at least 32 values per chunk.  The maximum amount of cache required for a column with 1 billion rows without a repetition index would be ~1.28 GiB.  In both Parquet and Lance, these search caches could likely be compressed (this would be trivial in Lance as the on-disk size of the miniblock metadata is already 2 bytes instead of 24 bytes).

\subsection{Struct Packing} \label{struct-packing}

The last technique we explore is \textbf{struct packing}.  When a struct is packed, we store the entire struct as one column instead of splitting the struct into leaf columns.  This increases the random access throughput since it allows us to retrieve multiple columns in fewer IOPS.  In exchange, we lose the ability to project columns from the struct.  We must fetch the entire struct and discard the data we are not interested in.

One significant detail in struct packing is that we first compress each column individually and then pack the buffers after the fact.  This gives us vectorized columnar compression.  In order for the packed struct to have fixed width, each field must be fixed width.  If any of the columns in the struct have variable width, then the entire struct is considered a variable width column.  A final difficulty is that we must ensure each column is chunked at the same rate if the packed data type is small enough to be eligible for miniblock.

In the extreme case, we can pack the entire record.  This converts Lance into a row-based storage format.  This technique could be useful in certain situations such as out-of-core sorting and hash-join algorithms.  For example, the lack of a standard row-based format for Arrow data has led Arrow-native engines such as DataFusion to use their own ad-hoc row-based solutions.

\section{Methods}

To verify our understanding, we have run a series of experiments measuring the performance of different structural encoding schemes.  These experiments were run with a Samsung 970 EVO Plus 2TB NVMe drive and an Intel i7-10700K CPU with 8 cores (16 threads).  We have benchmarked the hardware and found the peak performance of the disk to be ~850K random reads per second (at 4KiB) and ~3,400MiB/s throughput.

Library support is crucial for efficient random access.  Publicly available high-level libraries for Parquet and Arrow often lack effective APIs for random access.  Benchmarking against the available APIs is not an accurate representation of the capabilities of the format itself.  For Parquet, especially, there is a wide gap between what the format can achieve and what users typically use.  For example, with the default configuration of parquet-rs, we achieve ~5,500 rows per second on small scalars.  With optimized configuration as described below, we are able to achieve ~350,000 rows per second.

\subsection{Lance} \label{lance-methods}

Experiments involving Lance used the \texttt{lance} Rust crate.  Datasets were created with default settings which write 1 Mi rows per file with a default disk page (i.e. column chunk) size of 8 MiB.  Experiments used the file reader and writer directly to avoid dataset / table overhead and allow for more direct comparisons.  All experiments used the default parameters for the reader and writer.

\subsection{Parquet}

Experiments involving Parquet files used the \texttt{parquet-rs} Rust crate.  We found this to be more tunable and perform better than any Parquet alternatives.  To perform random access, we utilized the row selectors mechanism.   We built our own wrapper which manages the multiple files and creates a thread task per row group.  We performed a number of experiments with different row group and thread combinations, but unless otherwise specified, we only report the combination that yielded the best results.

\subsection{Arrow}

We were unable to achieve good performance with the existing Arrow IPC reader and writer APIs.  To study the performance of Arrow's structural encodings, we used the 2.0 version of the Lance file format.  This format is close to the strategy described in section \ref{arrow-structural}.  The only difference is that the Lance 2.0 format uses special offsets to represent null lists instead of a dedicated validity bitmap.  The same defaults and APIs were used that we described in section \ref{lance-methods}.

\subsection{Coalesced Access}

When performing random access into a file we can sometimes benefit when two rows are close together.  If we are performing the two accesses at the same time (i.e. they are part of the same operation) then we can combine the two I/O requests into a single IOP.  This reduces both the number of IOPS and the number of system calls.  If the two requests are close together in time but not part of the same operation then we can still benefit because the first request will likely still be in the OS page cache.  This reduces the number of IOPS but not the number of system calls.

\begin{figure}[h]
    \centering
    \includegraphics[width=8cm]{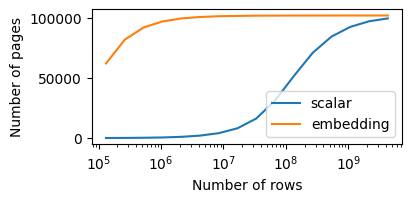}
    \Description{This chart shows that the benefits of coalesced access evaporate at four billion rows for 4 byte scalars and at one hundred thousand rows for 3 kilobyte embeddings.}
    \caption{The benefits of coalesced access diminish with dataset size}
    \label{fig:coalesce}
\end{figure}

We refer to both phenomena as \textit{coalesced access}.  We find that the benefits of coalesced access are easily overstated in benchmarking and less frequent in real-world use cases.  This is because the odds of benefiting from coalescing are dependent on the data type, the number of rows in the dataset, and the compression ratio of the data.  Figure \ref{fig:coalesce} demonstrates this phenomenon.  This chart shows the number of 8KiB pages selected if we randomly select 100,000 rows.  In smaller datasets, the odds of overlap are higher, and the benefit of coalesced access is high.  However, by the time our dataset has billions of rows, we are unlikely to see coalesced access benefits even for very small scalars.

The second problem with coalesced access is that it depends on the amount of page cache available.  In benchmarking, the system is normally isolated, and there is a large amount of page cache.  In production, we are intermixing small data type and large data type access, and we are reserving large blocks of memory for our search cache and possibly also secondary index storage.

For these reasons, we have designed our random access experiments to run on large datasets and random data.  We use one billion rows for small types.  This strikes a balance where there are still some benefits from coalesced access, but it is not overwhelming.  With larger data types, we see very little benefit from coalesced access at any scale we test at, and we are able to work with smaller datasets for expediency.  For exact details, refer to the scripts provided to reproduce our results.

\section{Experiments}

In this section we present a number of experiments to validate the benefits and drawbacks of different structural encodings that we presented in section \ref{strategies}.  In sections \ref{random-access-experiments} and \ref{full-scan-experiments} we look at the effect structural encoding has on random access and full scan workloads.  In section \ref{compression} we verify that compression is still achievable with the encoding strategies introduced.  Finally, section \ref{struct-packing-experiments} explores the effects of column packing on random access.

\subsection{Random Access} \label{random-access-experiments}

Random access patterns, often referred to as \textit{point lookups}, are important for many search-oriented use cases, as described in section \ref{introduction}.  Initially, we examine the overall random access performance of different approaches across a variety of data types that we have encountered in production applications.  These data types were intentionally chosen to explore the different structural encoding strategies presented in section \ref{strategies}.  For these experiments, we use random data.  All arrays contained a small portion (10\%) of null values.  For this experiment, only the top-level data type contained null values.

\begin{center}
\begin{tabular}{|c c c c|}
\hline
Name & Data Type & Repetition & Avg. Size Bytes \\
\hline
\hline
Scalar & UInt64 & No/Fix & 8 \\
\hline
String & Utf8 & No/Var & 16 \\
\hline
Scalar-list & List<UInt64> & Yes/Fix & 40 \\
\hline
String-list & List<Utf8> & Yes/Var & 80 \\
\hline
Vector & FSL<Float32,768> & No/Fix & 3Ki \\
\hline
Vector-list & List<FSL<Float32,768> & Yes/Fix & 15Ki \\
\hline
Image & Binary & No/Var & 20Ki \\
\hline
Image-list & List<Binary> Yes & Yes/Var & 100Ki \\
\hline
\end{tabular}
\end{center}

In this first experiment, we performed a number of "take" operations in parallel.  Each take operation selected 256 random indices from a single column of data.  We ran this workload for 10 seconds and recorded the average number of rows fetched per second.  We experimented with other take sizes as well; generally, there are some benefits to larger takes, as some overhead can be amortized, but this was small and consistent across formats and encodings.

We ensured the dataset is large enough so that indices are never repeated.  The scale was also large enough that benefits from coalesced I/O were unlikely.  This mirrors the search needs we have encountered where users typically have billions of rows and secondary indices are not correlated with the row order.  In these experiments the \textit{baseline} refers to the maximum our disk is capable of, without any benefit of coalesced I/O, as measured by independent benchmarks.

\subsubsection{Parquet}

\begin{figure}[h]
    \centering
    \includegraphics[width=6cm]{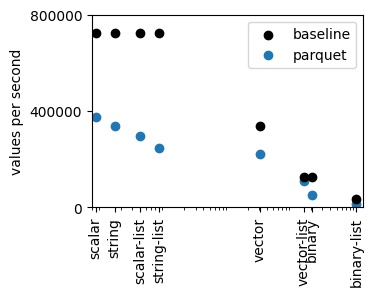}
    \Description{This chart shows that Parquet is able to read hundreds of thousands of rows per second across all take sizes and most data types.}
    \includegraphics[width=8cm]{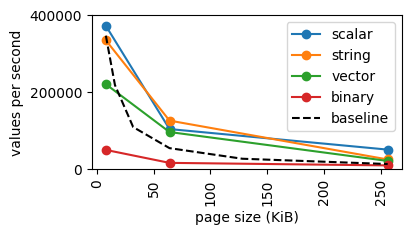}
    \Description{This chart shows that uncompressed and compressed data have little difference in takes per second but dictionary encoded data has terrible performance.}
    \caption{Parquet Random Access Performance}
    \label{fig:parquet-random}
\end{figure}

We first examine the random access performance of Parquet. In figure \ref{fig:parquet-random} we found that Parquet performed strongly across smaller data types.  Parquet's structural encoding requires a single IOP per row to fetch the page the row belongs to.  All buffers making up the array will be contained in that single IOP.  In this experiment, the page size was set to 8KiB and so each random access required a single 8KiB IOP.  There is room for improvement with smaller data types, and we discuss this further in section \ref{small-access-perf}.  As the data types became larger, Parquet's performance began to match the baseline.

As expected, the Parquet encoding scheme is highly dependent on the page size.  The second chart in figure \ref{fig:parquet-random} shows the correlation between page size and performance.  This is unsurprising, as our disk benchmarks confirmed that the size of an IOP quickly becomes the determining factor in random access performance on NVMe storage.

The exact relationship is not as direct as might be expected because in many cases we are able to exceed the IOPS rate of the disk due to the benefits of coalesced access, since these benefits increase as the page size increases.  The dotted line represents the peak IOPS the disk can support.  With 8KiB pages we are only able to exceed the disk rate with scalars.  At 64KiB pages we can exceed the disk rate with most data types.  However, the benefits we gain from coalesced access do not make up for the cost of the larger IOPS and performance still quickly declines.

We performed these experiments with compression disabled and dictionary encoding turned off, as this yielded the best performance.  Compression had fairly minor impacts on performance (88\% of ideal).  Dictionary encoding had a much more significant impact on performance (2\% of ideal).  This was surprising because we were working with random data, and we did not expect it to be dictionary encoded in the first place.  Parquet readers could likely overcome this limitation, without changes to the format, by limiting dictionary encoding to incoming arrays with low cardinality and by including dictionary pages as part of the search cache, similar to Lance.

\subsubsection{Arrow}

\begin{figure}[h]
    \centering
    \includegraphics[width=6cm]{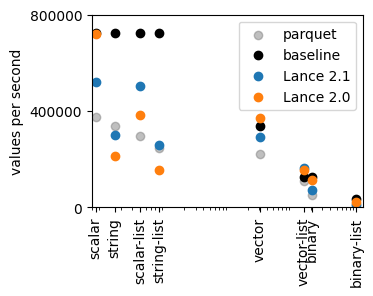}
    \Description{This chart shows that the Lance 2.0 format had better random access performance on scalar and vector types but suffered on string and string-list types}
    \includegraphics[width=8cm]{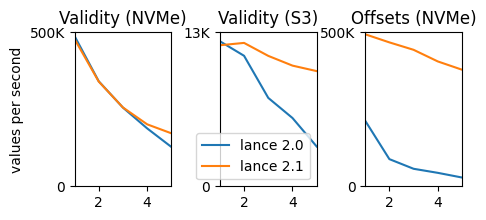}
    \Description{This chart shows the impacts of nesting level on performance.  The line for Lance 2.0 which represents Arrow style encodings, moves down and to the right.  The lines for Parquet and 2.1 have less of slope.}
    \caption{Random access performance of Arrow style encodings (via Lance 2.0) and the Lance encoding scheme (via Lance 2.1)}
    \label{fig:lance-random}
\end{figure}

Next, we ran the same experiment with Arrow structural encodings using the Lance 2.0 file format.  Figure \ref{fig:lance-random} shows the results, and we can see the performance exceeds Parquet and Lance 2.0 on simple types but is worse with types like string or string-list.  This is because the number of IOPS we must perform depends on the data type.  The extra IOPS required for string and string-list columns are substantial enough to hurt performance.  The extra IOPS required for fixed-width encodings (to load the validity) aren't as significant because coalesced access is giving us a lot of benefit (we are effectively treating the validity bitmaps as a part of our search cache).

To explore this further, we ran an additional experiment where we examined the performance on deeply nested data.  The second chart in figure \ref{fig:lance-random} shows the results.  Each level of nesting represents either an additional validity buffer or additional offsets buffer that we must fetch in addition to the values.  In the Lance 2.1 encodings, this does not represent an additional IOP.  In the Lance 2.0 format, which uses Arrow-style encodings, we do pay the cost of this extra IOP.

The effect is more significant in S3, where IOPS are far more expensive, or when dealing with list offsets.  When dealing with validity on NVMe, the effect was much less noticeable.  Validity buffers are small and benefit greatly from coalesced I/O.  We end up effectively adding the validity buffers to our search cache, which may be a valid strategy in many scenarios.

\subsubsection{Lance}

Finally, we examined the performance of the Lance 2.1 encoding scheme described in section \ref{lance-structural}.  The results of our random access experiment are also shown in figure \ref{fig:lance-random}.  We make up for our losses on string data and generally tie or beat Parquet; however, we slightly underperform the 2.0 format on simpler types.  As we see in the second chart, this format gains robustness against levels of nesting.

\begin{figure}[h]
    \centering
    \includegraphics[width=8cm]{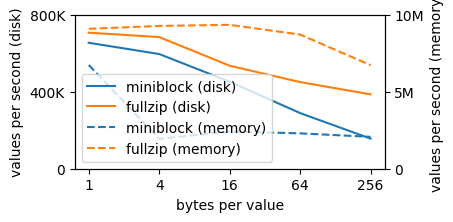}
    \Description{Full zip always has better random access performance.  This difference is even more considerable when the data is in memory.}
    \caption{Full zip encoding is lighter weight and significantly better in random access}
    \label{fig:lance-sized}
\end{figure}

One potential reason for underperforming on scalar data is that we must decode an entire page (miniblock chunk) of data in order to obtain a single value.  This overhead is required because we allow for the use of dense and opaque encodings to be used within the miniblock chunk.  To help understand this effect, we look at our performance on different data sizes in figure \ref{fig:lance-sized}.  The full-zip encoding is far more lightweight and performs better at all data sizes.  This difference is much greater when the data is in memory and no I/O is required.

\subsubsection{Profiling Small Access Performance}
\label{small-access-perf}

Both Parquet and Lance fell short of the NVMe's capabilities with small access.  From profiling, we have determined two likely root causes that will require future investigation.  First, no effort is made to align pages to disk sectors, and both formats required 2-3 disk sectors per I/O.  Second, both formats were using the pread64 system call for reads.  This introduces considerable system call overhead in addition to the decompression work that must be done.

It may be possible to reduce this work considerably using I/O approaches (such as io\_uring) that work with kernel buffers.  Also, these small data types all used the miniblock structural encoding which utilizes opaque encodings.  It would also be possible to reduce the decompression work by restricting ourselves to transparent encodings.  We leave this exploration for future work.

\subsection{Compression Analysis} \label{compression}

We next examine the compression effectiveness on our sample data types.  This paper does not aim to describe the ideal compression scheme or invent new compressive encodings.  However, we must show that we can still achieve effective compression and full scan performance with our chosen structural encodings.  In order to get measurements representative of real-world scenarios, we have found practical examples of data, focusing on use cases important to ML workloads and exploring a variety of compression scenarios.  The following table describes the example scenarios we have chosen and describes the compression used by Lance.  In Parquet, we used Snappy and enabled dictionary encoding for all full scan tests.

\begin{center}
\begin{tabular}{|c c c|}
\hline
Name & Data Type & Compression (Lance) \\
\hline
\hline
names & US baby names \cite{BabyNames} & Dictionary + FSST \\
\hline
prompts & LLM training prompts \cite{UltraChat} & FSST \\
\hline
dates & TPC-H ship date & Bitpacking \\
\hline
reviews & Amazon reviews \cite{AmazonReviews} & FSST \\
\hline
code & Source code \cite{GithubCode} & LZ4 \\
\hline
images & Compressed images \cite{TakaraImages} & LZ4 \\
\hline
embeddings & CLIP image embeddings \cite{Laion5b} & None \\
\hline
websites & HTML websites \cite{CommonCrawl} & LZ4 \\
\hline
\end{tabular}
\end{center}

\begin{figure}[h]
    \centering
    \includegraphics[width=8cm]{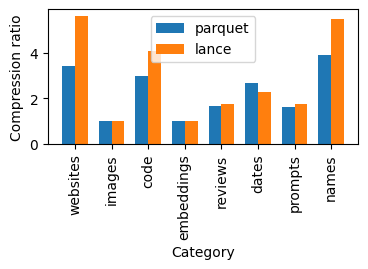}
    \Description{Shows the compression ratios of the different scenarios.}
    \caption{Lance's structural encodings can still compress as efficiently as Parquet}
    \label{fig:uncompression}
\end{figure}

Figure \ref{fig:uncompression} shows the compression ratios of each of our scenarios.

We only study the Parquet encoding scheme and Lance encoding scheme when discussing full scans and compression.  Both the Arrow IPC file format and ORC can perform compression on an Arrow-style structural encoding, but available implementations currently handle compression in a way that makes random access impossible, and so we do not consider them.

It would be possible to use Arrow-style structural encodings, with compression, and still achieve random access.  For example, by using techniques such as bit-packing, FSST, or per-value encodings which have no dependencies between values.  However, there is no reason to believe these approaches would be more effective (from a compression ratio perspective) than those used here.

\subsection{Full Scan} \label{full-scan-experiments}

To measure full scan performance, we found example datasets for each of our scenarios.  We then calculated the rate at which we can read the data.  We measure both the compressed (disk) throughput and the iterations per second.  The former helps us understand how effectively we are utilizing the disk, while the latter helps us compare true performance when comparing formats.  For example, an expensive but highly effective compression might lead to poor disk utilization but still deliver high performance.

\subsubsection{Parquet}

\begin{figure}[h]
    \centering
    \includegraphics[width=8cm]{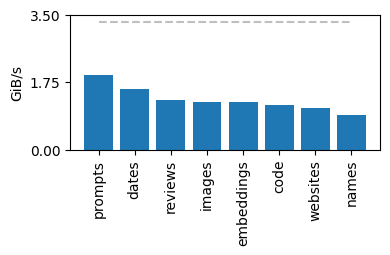}
    \Description{This chart shows scan performance versus page size.  Most lines are horizontal, indicating no significant effect.  The code category, on the other hand, increases significantly with page size.}
    \caption{Parquet does not fully utilize the available disk bandwidth during scans}
    \label{fig:parquet-scan-disk}
\end{figure}

In figure \ref{fig:parquet-scan-disk} we evaluate how effectively Parquet is able to utilize the NVMe during a full scan.  We tested page sizes of 8, 32, 64, and 256 KiB.  We also tested row group sizes of 1 Ki, 10 Ki, 100 Ki, and 1 Mi.  For each category, we only report the best score across all combinations of page size and row group size.  Parquet falls short of utilizing the disk's capabilities.  For many encodings, we only achieve half of the disk's peak bandwidth.

Our belief is that the reason for Parquet's ineffective disk utilization is not because Parquet is compute bound.  Instead, from our profiling, we believe the issue is ineffective I/O scheduling.  We can run many Parquet readers in parallel, but each reader is alternating between scheduling and waiting for I/O and decoding data.  At any moment, if all threads are busy decoding, then the disk is possibly sitting idle.  Increasing the number of threads does not prevent all cores from becoming busy.  

We also examined the effects of page size on scan performance.  We calculated a normalized performance score where 1 represented the best score across all page sizes and row groups for a category.  Surprisingly, 8KB page sizes scored the best in all categories but one (embeddings) and the normalized score for 8KB in that category was 0.997.  In other words, 8KB pages are ideally suited for both random access and full scan performance against NVMe.

In these experiments we disabled all page-level statistics.  These are extremely detrimental to both scan performance and file size when working with small pages.  We do not believe statistics should be stored inline at the page or the row group level.  Instead, they should be stored separately, and only loaded when required.  If a file format supports fast random access with minimal read amplification, then there is no need for statistics to be stored inline.  This is supported by \cite{DecomposeFileFormat}

\begin{figure}[h]
    \centering
    \includegraphics[width=8cm]{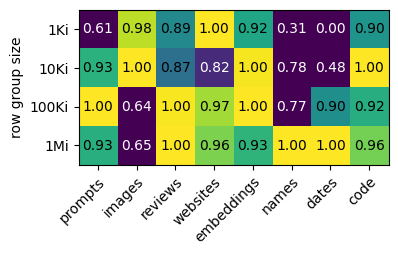}
    \Description{This heat map shows performance of different row group sizes and different categories.  Some categories prefer small row groups and others prefer large.  It almost seems random.}
    \caption{Row group size has significant effect on scans}
    \label{fig:parquet-row-group-size}
\end{figure}

Finally, we found configuring Parquet's row group size to be difficult.  Figure \ref{fig:parquet-row-group-size} demonstrates the effects of row group size on scan performance (the dates column failed with a row group size of 1Ki as this resulted in more than $2^{16}$ row groups).  Generally, smaller types prefer larger row groups, and larger types prefer smaller row groups.  However, files typically have many columns, and a single row group size must be chosen across the entire file.

\subsubsection{Lance}

\begin{figure}[h]
    \centering
    \includegraphics[width=8cm]{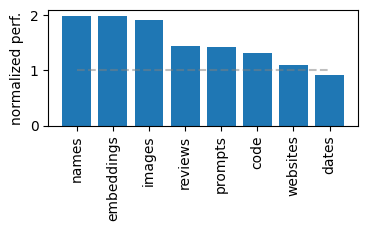}
    \Description{This chart shows Lance scan performance vs Parquet scan performance.  It is a bar chart where the y axis goes from 0 to 2.  The bars for names, embeddings, and images are in the 1.8-2.0 range.  The bars for reviews, prompts, and code are in the 1.3-1.5 range.  The bar for websites is slightly over 1.0.  The bar for dates is slightly below 1.0.}
    \includegraphics[width=8cm]{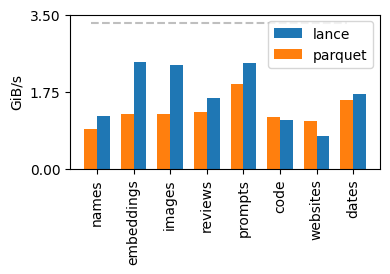}
    \Description{This chart shows how effectively Lance is able to use the disk.  It is a bar chart with two groups, Lance and Parquet.  The Y axis is disk utilization in GiB/s.  In images, embeddings, names, reviews, prompts, and dates Lance is higher than Parquet.  In code and websites Lance is lower than Parquet.  In embeddings, images, and prmopts Lance is significantly larger.  In most cases all bars are still well under what the disk is capable of.}
    \caption{Lance scan performance often exceeds Parquet due to more efficient disk utilization}
    \label{fig:lance-vs-parquet-scan}
\end{figure}

In figure \ref{fig:lance-vs-parquet-scan} we repeat the same scan experiment with Lance style structural encodings.  Our score in this figure is a normalized performance score.  A value of 1 represents the best performance (across all page sizes) that we obtained with Parquet for that category.  Performance in this case is measured in file reads per second and not based on disk throughput.  This ensures that compression effectiveness plays a role and we are not unfairly benefiting from lightweight compression techniques.

\begin{figure}[h]
    \centering
    \includegraphics[width=8cm]{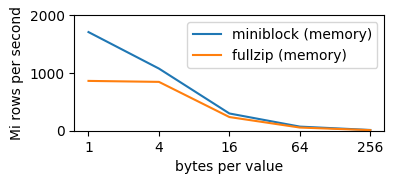}
    \Description{There are two pairs of lines, one pair for disk and one for memory.  Each pair represents full zip and miniblock encoding.  The disk lines never cross, full zip is always superior.  The memory lines cross, below 64 bytes per value the miniblock encoding gets better performance.}
    \caption{The miniblock encoding does less work during full scans since there is less CPU work per-value}
    \label{fig:sized-scan}
\end{figure}

We are able to achieve full scan performance equal to or greater than Parquet in nearly all cases.  Part of this reason is that the Lance encoding scheme is more able to exploit the disk as we see in the second chart in figure \ref{fig:lance-vs-parquet-scan}.  The Lance structural encodings allow us to fully exploit the disk in both random access and full scan scenarios with a single configuration for all data types.  We note that there is still significant room for improvement, in particular when utilizing LZ4 compression.  In some cases, we suspect the Lance format is CPU bound on decompression and there is room for compression optimization.  In others, such as names and dates, we believe it could more efficiently schedule I/O.

We can also see in figure \ref{fig:sized-scan} how the different structural encodings support full scans.  When we are not I/O bound, there is a significant advantage to using the miniblock encoding.  This is because the unzipping process becomes expensive as it needs to be applied on a per-value basis without any vectorization.  This is one of the reasons we use miniblock encoding for small data types.

\subsection{Struct Packing} \label{struct-packing-experiments}

The final experiment we look at measures the effectiveness of struct packing.  This was described in section \ref{struct-packing}.  By packing multiple fields together, we should be able to improve our random access performance when accessing both fields.  However, our full scan performance should drop when accessing one of the fields in isolation.

\begin{figure}[h]
    \centering
    \includegraphics[width=8cm]{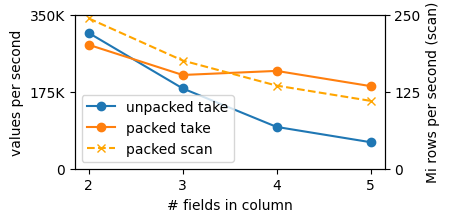}
    \Description{This chart shows the effect of packing.  There is a line for unpacked and packed.  Both lines decrease performance as the number of fields grow but the slope for packed is much gentler.  By the time we have five fields the packed approach has 2x the performance of the unpacked approach.}
    \caption{Packing structs trade single field scan speed for whole struct random access speed}
    \label{fig:packed-take}
\end{figure}

To measure this we evaluate structs with 2, 3, 4, and 5 fields.  We look at the random access performance when accessing all fields in the struct as well as the full scan performance when accessing a single field in the struct.  This experiment does not use any compression and so the cost for loading a single field is the same as the cost for loading the entire struct.  This experiment used random data and small scalar fields.  The results are shown in figure \ref{fig:packed-take}.  As expected, when using packed structs our random access performance falls more slowly while our full scan performance drops in an approximately linear fashion.

\section{Conclusion}

We have introduced the concept of structural encoding and used it to evaluate existing columnar formats.  We have described shortcomings, challenges, and improvement opportunities for structural encodings used by several common file formats.  We introduced the Lance structural encoding scheme, which alternates between two different structural encodings to balance performance on small and large data types.

We have developed a fully functional file reader and writer using this scheme.  We have shown that Parquet and Lance style encodings are able to achieve solid performance in both full scans and random access.  However, Lance style encodings allow for a smaller search cache, avoid the need to configure row group sizes, and enable additional capabilities such as struct packing, bridging lists across pages, and nested repetition indices.  We have shown that Arrow style encodings require too many IOPS and cannot achieve sufficient random access against nested data types, especially against cloud storage.  Both the Parquet and Lance schemes are able to effectively utilize NVMe storage, although both have room for improvement.

\section{Future Work}

In this paper, the concept of structural encoding helps to explore trade-offs in storage, compute, and RAM.  We believe there may be different structural encoding schemes that could be utilized, either to achieve better performance in NVMe storage, to be applied to different storage situations (in-memory, NVMe array, S3), and to work with new data types or encodings (such as variant encodings).  For example, using Lance as row-storage by packing the top-level struct is an unexplored technique and would be interesting to compare with common row-storage formats.

We believe this research also has a significant impact on those developing new file formats, libraries, and standards, which may someday replace the columnar formats we have today.  We believe there is no universally correct structural encoding, and we hope that these new formats will avoid "baking in" a single structural encoding and will aim to make this a configurable strategy.  This can help avoid a proliferation of single-purpose formats in favor of configurable multi-purpose formats.

\begin{acks}
 We would like to extend special thanks to the authors' families, who kept our spirits up during long hours of research, testing, and countless unexpected results that required further questioning.  We'd also like to thank the many users of Lance and LanceDB whose feedback, guidance, suggestions, and testing got us to where we are today.
\end{acks}


\balance
\bibliographystyle{ACM-Reference-Format}
\bibliography{main}

\end{document}